\documentstyle[nato,fps]{crckapb} 

\def\lsi{\raise0.3ex\hbox{$<$\kern-0.75em\raise-1.1ex\hbox{$\sim$}}}
\def\gsi{\raise0.3ex\hbox{$>$\kern-0.75em\raise-1.1ex\hbox{$\sim$}}}

\newcommand{\gsim}{\mathop{\gsi}}

\begin{opening}
\title{Optimizing Chirality and Scaling of Lattice Fermions}

\author{W. Bietenholz}
\institute{NORDITA, \ \
Blegdamsvej 17 \\
DK-2100 K\o benhavn \O , Denmark \\
{\bf NORDITA-2000-1-HE, hep-lat/0001001}}

\end{opening}

\runningtitle{Optimizing Chirality and Scaling of Lattice Fermions}
\vspace{-5mm}

\begin{document}


\vspace{-5mm}
\section{Introduction}

If we construct a lattice fermion formulation, there
are a number of goals to be considered:
{\em doubling} should be avoided;
even at finite lattice spacing $a$, we want to represent 
{\em chiral symmetry} in a sound way; and we
are seeking a good {\em scaling} behavior. Conceptually
we have to require {\em locality} (the lattice Dirac operator $D(x,y,U)$
has to decay at least exponentially in $\vert x-y \vert$ ).
In addition, for practical purposes we desire a high level 
of locality, i.e.\ a {\em fast} exponential decay or even 
ultralocality (which means that the couplings in $D$ drop to zero
beyond a finite number of lattice spacings).
A further issue is a good approximation to rotation invariance.
Last but not least, the formulation should be simple
enough to allow for efficient simulations.
Here we report on a construction, which is designed to
do justice to all of these goals.

\section{Ginsparg-Wilson fermions (an unconventional introduction)}

For a lattice Dirac operator $D$, full chiral invariance
( $\{ D , \gamma_{5} \} = 0$ )
is incompatible with other basic requirements (Hermiticity,
locality, absence of doublers, discrete translation invariance) 
\cite{NN}. Therefore we only implement a
{\em modified chiral symmetry}, which does allow $D$ to fulfill
those requirements. For such a modified
chiral transformation we start from the ansatz
\begin{equation}
\bar \psi \to \bar \psi \, ( 1 + \epsilon [ 1 - F] \gamma_{5} ) \ , \
\psi \to ( 1 + \epsilon \gamma_{5} [ 1 - G] ) \, \psi \ ,
\end{equation}
$\epsilon$ being an infinitesimal transformation parameter.
The transformation, and therefore $F$ and $G$ should be local,
and $F, G = O(a)$, so that we reproduce the full chiral
symmetry in the (naive) continuum limit. 
\footnote{For convenience, we set $a=1$ in the formulae
(on an isotropic Euclidean lattice),
but we classify the terms nevertheless by the order of $a$
that they (would) belong to.}
Invariance of the Lagrangian
$\bar \psi D \psi$ holds to $O(\epsilon )$ if
\footnote{In our short-hand notation, the 'products'
are convolutions in c-space.}
\begin{equation} \label{trafo}
\{ D , \gamma_{5} \} = F \gamma_{5} D + D \gamma_{5} G .
\end{equation} 
This implies a continuous modified chiral symmetry, which
has the full number of generators. It may be compared
to the remnant chiral symmetry of staggered fermions:
there the doubling problem is not solved, and
one is only left with a $U(1)\otimes U(1)$ symmetry, which does,
however, protect the mass from additive renormalization.
The same can be shown here if we assume ``$\gamma_{5}$-Hermiticity'',
$D^{\dagger} = \gamma_{5} D \gamma_{5}$, and we choose
$F = D R$, $G = R D$,  where $R$ is {\em local} again, non-trivial
and $[R,\gamma_{5}]=0$ (this generalizes Ref.\
\cite{ML}). Then eq.\ (\ref{trafo}) turns into the
{\em Ginsparg-Wilson relation} (GWR) \cite{GW}
\footnote{$\gamma_{5}$-Hermiticity is essentially inevitable for any
sensible solution, but if we want to formulate the GWR even without
this assumption, then it reads $\{ D,\gamma_{5} \} = 2 D R \gamma_{5}D$.
This follows from the immediately obvious prescription
$\{ D^{-1},\gamma_{5} \} = 2R\gamma_{5}$. Alternatively, if we require
$D$ to be normal we arrive at $\{ D,\gamma_{5} \} = 2 R D \gamma_{5}D$ 
\cite{Kerl}.
However, for the results presented in Sec. 3, 4 this doesn't 
matter, since we always use $R_{x,y} \propto \delta_{x,y}$.}
\begin{equation} \label{GWR}
D + D^{\dagger} = 2 D^{\dagger} R D \ ,
\end{equation}
and it implies the absence of additive mass renormalization 
(see also Ref.\ \cite{Has}) since
\begin{equation} \label{spec-con}
( \sqrt{2R} D^{\dagger} \sqrt{2R} - 1) ( \sqrt{2R} D \sqrt{2R} - 1) = 1 \ .
\end{equation}
As another illustration we can write the GWR as
$\{ D^{-1},\gamma_{5} \} = 2R \gamma_{5}$, and we see that a local term $R$
does not shift the poles in $D^{-1}$ (in contrast to the cases
where $\{ D,\gamma_{5} \} /2$ is local, such as a mass
or a Wilson term).
\footnote{An analogous treatment is also conceivable in the continuum.
For comments related to the Pauli-Villars regularization, see Ref.\ \cite{Fuji}.
In dimensional regularization $\gamma_{5}$ is a notorious trouble-maker.
It may be useful to substitute it by operators
$(1-F)\gamma_{5}$ resp.\ $\gamma_{5}(1-G)$ at the suitable places.
In $d+\varepsilon$ dimensions ($d$ even), $F$ and $G$
could take the form $\varepsilon DR/\mu$ resp.  $\varepsilon RD/\mu$
(where $R$ is some local term, and $\mu$ is the usual scale in dimensional regularization).
We expect the chiral anomaly to be reproduced correctly as $\varepsilon \to 0$.}
[With respect to the general ansatz, we have to require the right-hand
side of $ D^{-1} + D^{\dagger~-1} = D^{\dagger~-1}\gamma_{5} F \gamma_{5} 
+ GD^{-1} =  D^{-1}F + \gamma_{5} G \gamma_{5} D^{\dagger~-1}$ to be local.]

$\gamma_{5}$-Hermiticity implies $R^{\dagger}=R$.
If we now start from some lattice Dirac operator $D_{0}$
(obeying the assumption of the Nielsen-Ninomiya theorem 
such as absence of doublers, but otherwise quite arbitrary),
we can construct a Ginsparg-Wilson operator $D$ from
it by enforcing eq.\ (\ref{spec-con}) as
\begin{equation} \label{overlap}
D = \frac{1}{\sqrt{2R}} \Big[ 1 + \frac{A}{\sqrt{A^{\dagger}A}} \Big]
\frac{1}{\sqrt{2R}} \ , \quad A := \sqrt{2R} D_{0} \sqrt{2R} - 1 \ .
\end{equation}
This is the generalization \cite{EPJC} of the {\em ``overlap formula''},
which uses the {\em ``standard GW kernel''}
\begin{equation} \label{Rst}
R^{(st)}_{x,y} := \frac{1}{2} \delta_{x,y} \ , 
\end{equation}
and which leads from the
Wilson fermion $D_{0}=D_{W}$ to the Neuberger fermion $D=D_{Ne}$ \cite{Neu}.
In the general solution of the GWR, eq.\ (\ref{overlap}), we can obviously
vary the parameters in $D_{0}$ or in $R$ (or in both) in many ways,
without violating their required properties. This shows that there
exists a continuous set of GWR solutions $D$ in the space of coupling parameters.

Any solution of the GWR is related to a fully chirally invariant
Dirac operator $D_{\chi} = D (1 - RD)^{-1}$, which is, however,
non-local (in the free case, $D_{\chi}(p)$ has poles, cf.\ eq.\ (\ref{spec-con})). 
Vice versa, if we start from some $D_{\chi}$ with this type of non-locality
(such as the Rebbi fermion \cite{Rebbi}, for example) we can construct
a GW solution \cite{EPJC,CZ} $D = D_{\chi} (1+RD_{\chi})^{-1}$, which is local, 
at least in the free and weakly interacting case.
The mechanism of providing locality by inserting a local term $R \neq 0$
is known from the framework of perfect actions, where the factor $R^{-1}$
occurs in a Gaussian block variable renormalization group
transformation term of the fermions \cite{GW,UJW}.
Hence $R \to 0$ corresponds to a $\delta$ function block variable transformation, 
and the corresponding perfect action has a Rebbi-type non-locality. The transition
to locality requires the superficial breaking of the full chiral symmetry, $R\neq 0$:
chirality is manifest in the action only in the sense of the GWR, but it
is fully present in the physical observables \cite{Melb,BWNP}.
\footnote{Such a superficial symmetry breaking in the
transformation term is {\em not} necessary in order to preserve
{\em supersymmetry} in a RGT, an hence in a perfect action \cite{SUSY}.
A.\ Thimm also commented on an analogous treatment of further symmetries
\cite{Axel}.}
In contrast to the Rebbi fermion \cite{Peli}, the axial anomaly 
is correctly reproduced \cite{BWPL} for the perfect action at any
local term $R$, including the perfect $D_{\chi}$ (for $R=0$).
This should also be checked if one generally wants to use $D_{\chi}$
in an indirect way \cite{Capi}, by measuring the right-hand side of
$\langle D^{-1}_{\chi} \rangle = \langle D^{-1} \rangle - 
\langle R \rangle$.

By introducing a non-trivial kernel $R$ we have relaxed the condition
of chiral symmetry somewhat --- without doing harm to the physical properties
related to chirality \cite{GW,Has,ML} --- and this allows for locality of
$D$ (as well as the absence of doublers etc.)
\footnote{This hold at least as long as the gauge background is smooth.
At very strong coupling, locality is uncertain, and also the doubling
problem can return, see Subsec.\ 4.2.}
without contradiction to the Nielsen-Ninomiya theorem.
In the case of the Neuberger fermion $D_{Ne}$, locality has been
demonstrated in a smooth gauge background. In particular, zero
eigenvalues in $A^{\dagger}A$
are excluded if the inequality (in $d$ dimensions) 
\begin{equation}
\Vert 1 - P \Vert < \frac{b}{d(d-1)}
\end{equation} 
holds for any plaquette variable $P$ and a suitable bound $b$. 
From Ref.\ \cite{HJL} we obtain $b=0.4$, which has recently
been improved to $b= (1 + 1/\sqrt{2})^{-1} \simeq 0.586$ \cite{Neuloc}. 
Still, this constraint is somewhat inconvenient; for instance, 
at least one eigenvalue of $A^{\dagger}A$ has to cross zero
if we want to change the topological sector.

Furthermore, the GWR allows for locality only in the sense
that the couplings in $D(x,y,U)$ decay exponentially, 
but {\em not} for ultralocality.
This was conjectured intuitively in Ref.\ \cite{EPJC}. 
To demonstrate this No-Go rule for GW fermions,
it is sufficient to show it for the free fermion.
A proof, which was specifically restricted to $R =  R^{(st)}$,
has been given in Ref.\ \cite{IH}. By now, a complete proof
covering {\em all} (local) GW kernels $R$
has been added, hence this rule is completely general \cite{ultraloc}.

In that context, it is amusing to reconsider the ordinary Wilson fermion.
From the mass shift we know that it is certainly not a GW fermion in general,
and according to our No-Go rule not even the free Wilson fermions
can obey any GWR. Indeed, if we insert the free $D_{W}$ into
the GWR and solve for $R_{x,y}$, we find that it decays
$\propto \vert x-y \vert ^{-4}$ in $d=2$, 
and $\propto \vert x-y \vert ^{-6}$ in $d=4$,
which is nonlocal and therefore not a GW kernel.

The exponential decay of $D$ is satisfactory from the conceptual point of view,
but the existence of couplings over an infinite range 
is a problem for practical purposes. One would hope for
a {\em high degree of locality} at least, i.e.\ for a fast exponential decay.
Almost all the literature on overlap fermions solely deals with the
Neuberger fermion, but it turns out that the couplings in 
$D_{Ne}$ do not decay as fast as one would wish, see Fig.\ \ref{local}. 
Moreover, also other properties listed in Sec.\ 1 ---
most importantly scaling, but also approximate rotational invariance ---
are unfortunately rather poor. This is obvious even from the free fermion,
see Figs.\ \ref{freescale}, \ref{local}. 
On the level of the action, there are generally
no $O(a)$ artifacts for GW fermions, because any additional clover term
violates the GWR \cite{FN}, but from the free case we see
already that the $O(a^{2})$ scaling artifacts in $D_{Ne}$ are large.
\footnote{Our study of the 2d and 4d free fermion, as well as the
2d interaction fermion, show consistently that the artifacts in
$D_{Ne}$ are even worse than those in $D_{W}$. Ref.\ \cite{KFL}
is more optimistic about the scaling of the Neuberger fermion
(but it does not include a comparison to other types of lattice 
fermions).} And yet its simulation is tedious (the quenched case
requires already a similar effort as simulating $D_{W}$ with dynamical
fermions \cite{JanWup}), allowing only for the use of small lattices.

However, $D_{Ne}$ arises only from a very special choice
in a large class of GWR solutions described by eq.\ (\ref{overlap}),
namely $D_{0}=D_{W}$ and $R=R^{(st)}$. The message
of this report is that there are better options,
and we are going to show in Sec. 3, 4 how {\em improved
overlap fermions} can be constructed, tested and applied.
This generalization and the improvement concept
for overlap fermions was introduced in Ref.\ \cite{EPJC}. 
It was extensively tested in the framework of the 
Schwinger model \cite{WBIH}.

\subsection{THE CONCEPT OF IMPROVING OVERLAP FERMIONS}

We first summarize the main idea: if $D_{0}$ happens to
be a GW operator already (with respect to a fixed term
$R$), then eq.\ (\ref{overlap}) yields $D=D_{0}$; the operator
reproduces itself. Therefore, any GW fermion --- such as the
perfect \cite{GW} or the classically perfect \cite{Has} fermion ---
is automatically an overlap fermion too.
The construction of a classically perfect action
for asymptotically free models
only requires minimization --- no (numeric) functional integral,
in contrast to the perfect action --- and it still has
the additional virtues of excellent scaling and rotation 
invariance. Unfortunately, a really powerful quasi-perfect
action is not available so far for interacting fermions in $d=4$.

However, if we manage to construct at least an {\em approximate GW fermion},
then we can expect it to {\em change only modestly} if we insert it as
$D_{0}$ in the overlap formula (for the corresponding $R$), 
$D \approx D_{0}$. 
If our approximate GW fermion is in addition short-ranged, 
then we can expect $D$ to have a high degree of locality,
since the long distance couplings are turned on just a little
in $D$. (Also $D_{0}=D_{W}$ is short-ranged, but since
this is far from a GW fermion, it changes a lot in the overlap
formula, and those long distance couplings cannot be predicted
to be tiny). Similarly, if $D_{0}$ scales well, then
we can expect this quality to be essentially preserved
in $D$ if $D \approx D_{0}$, and the same argument applies
to the approximate rotation invariance. 
In Sec.\ 3 we discuss examples for
promising approximate GW operators. In Sec.\ 4 they are
transformed into exact GW fermions, and the above predictions
are verified.

\section{Short-ranged approximate Ginsparg-Wilson fermions}

\begin{figure}[htb]
\vspace{-3mm}
\begin{center}
\hspace{15mm}
\def\fpsangle{0}
\epsfxsize=80mm
\fpsbox{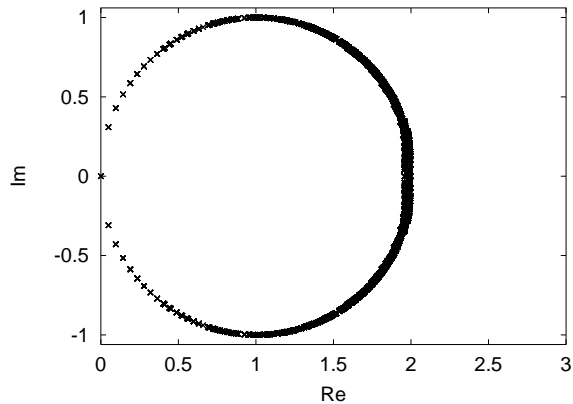}
\end{center}
\vspace{-5mm}
\caption{\it{The spectrum of a truncated perfect, free HF on a $20^4$ lattice
(plotted in $C \!\!\!\! \tiny{I}$).}}
\label{spec4d}
\vspace{-3mm}
\end{figure}

Perfect free fermions can be constructed and parameterized in
c-space explicitly \cite{BWNP}. If we choose
the term $R$ such that the locality is optimal at mass zero, then we arrive
at the standard form $R^{(st)}$. To make such a fermion tractable,
first of all its couplings have to be truncated to a short range.
A truncation to couplings inside a unit hypercube 
({\em ``hypercube fermion''}, HF) was performed in Ref.\ \cite{BBCW}
by means of periodic boundary conditions.
(Alternative 4d truncated perfect HFs can be found in
Refs.\ \cite{AHF}.)
Truncation causes some scaling artifacts, but they are small, so that
the free HF is still strongly improved over the Wilson fermion.
This can be observed from the dispersion relation as well as thermodynamic
scaling ratios \cite{BBCW,chempot}. At the same time, truncation also
implies a (small) violation of the GWR, in agreement with the absence
of ultralocal GW fermions. The spectrum of a GW fermion with
$R_{x,y} = \delta_{x,y}/(2\mu )$, ($\mu > 0$) is situated on a circle in
$C \!\!\!\! \tiny{I}$ with center and radius $\mu$ (GW circle), 
see eq.\ (\ref{spec-con}).
Hence we can test the quality of our approximate GW fermion
(with respect to $R^{(st)}$)
by checking how close its spectrum comes to a unit circle.
This is shown in Fig.\ \ref{spec4d} for the 4d HF of Ref.\ \cite{BBCW}
on a $20^{4}$ lattice. We see that we have a good approximation,
especially in the physically important regime of eigenvalues close to 0.

\begin{figure}[hbt]
\begin{tabular}{cc}
\def\fpsangle{270}
\epsfxsize=43mm
\fpsbox{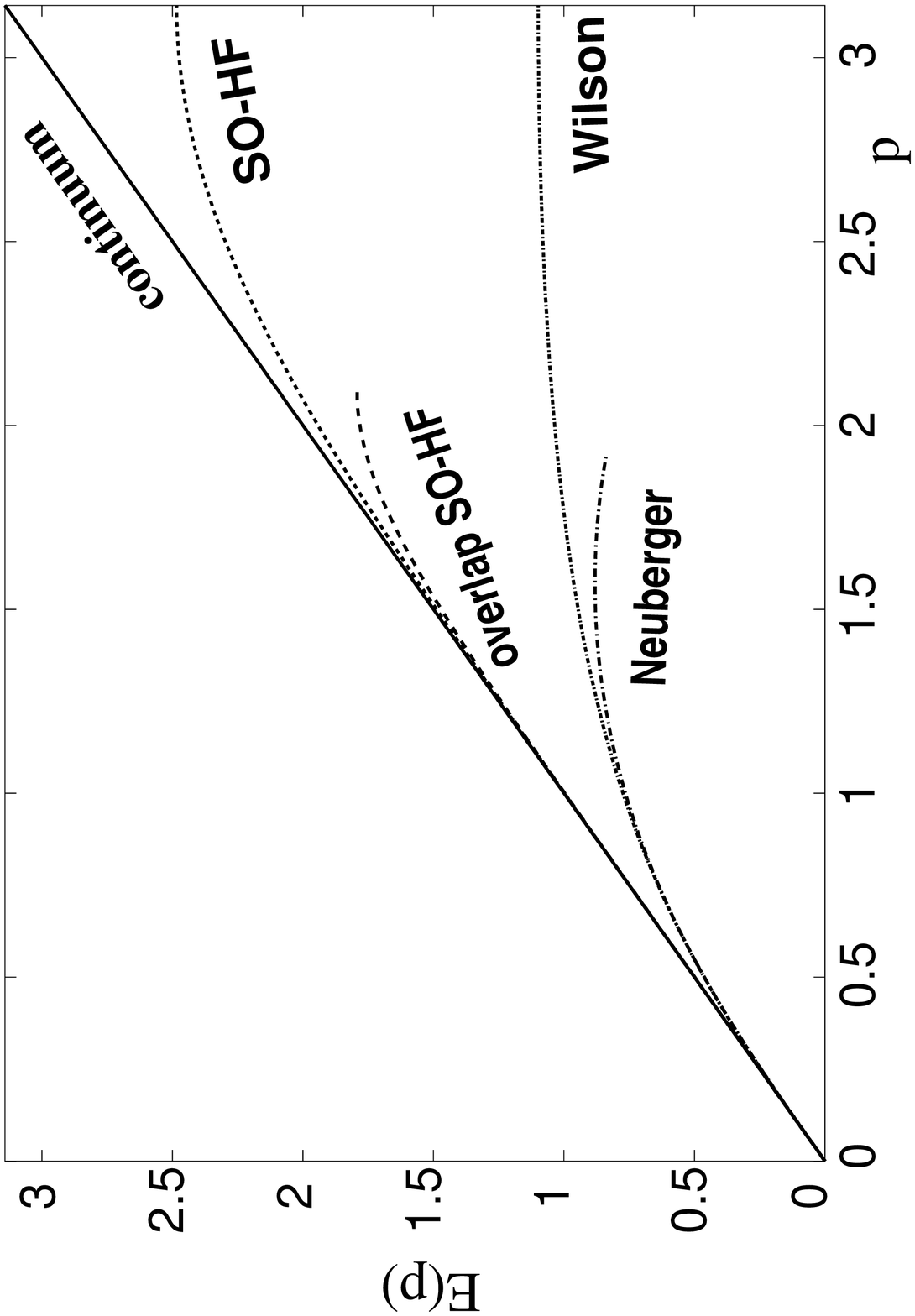}
\def\fpsangle{270}
\epsfxsize=43mm
\fpsbox{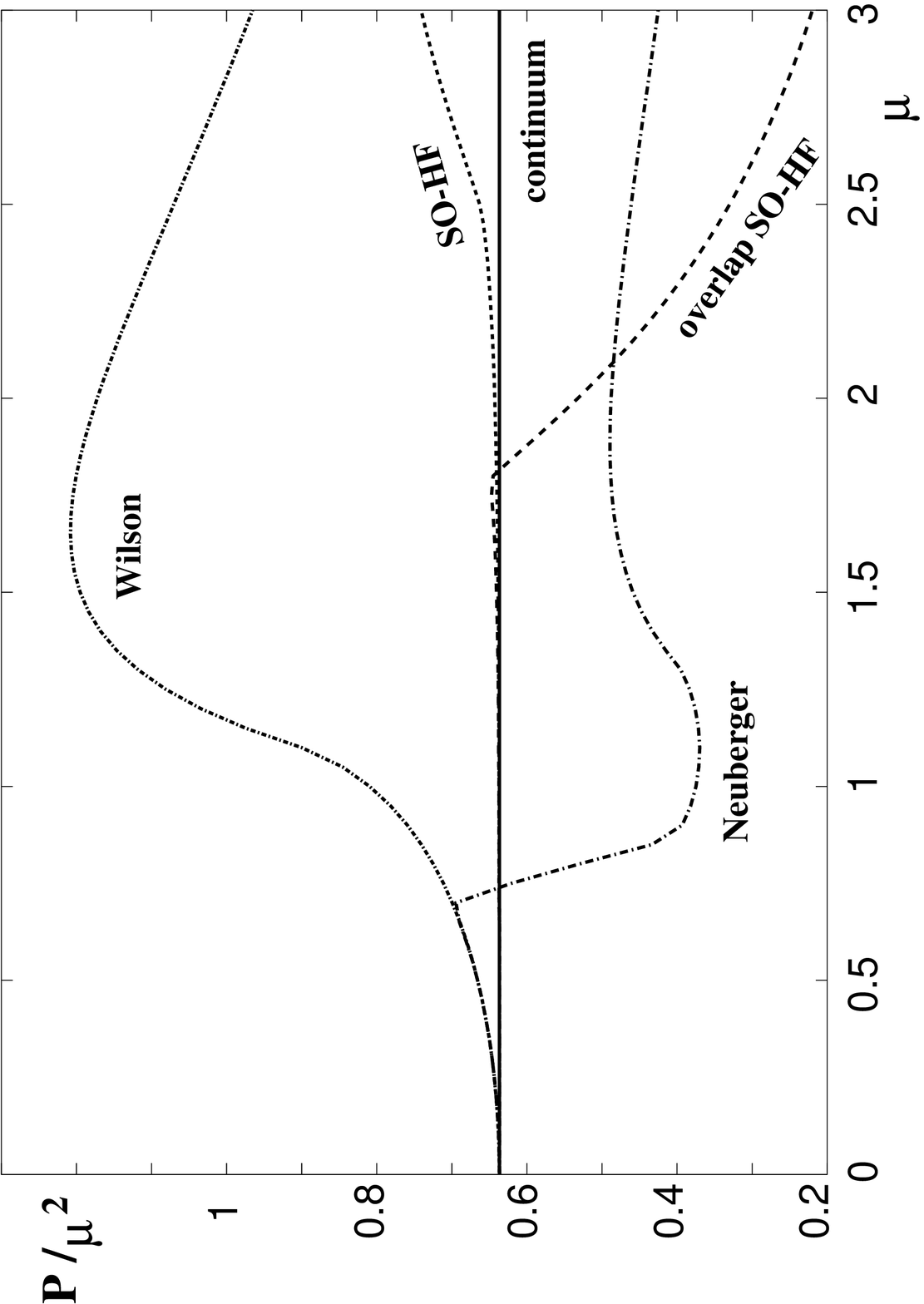}
\end{tabular}
\vspace{-3mm}
\caption{\it{Scaling of free fermions: the fermion dispersion relation
(left) and the thermodynamic scaling ratio \ $pressure/(chemical \ potential)^{2}$
at zero temperature (right).}}
\vspace{-3mm}
\label{freescale}
\end{figure}

In $d=2$ we start off from a similar massless HF, which is optimized for its
scaling behavior; its set of couplings is given in Ref.\ \cite{WBIH},
and its strong improvement over the Wilson fermion
is visible in Fig.\ \ref{freescale}.
\footnote{As an alternative, one could directly optimized the chiral
properties by minimizing the violation of the GWR \cite{EPJC,WBIH,GatHip}.
However, those properties can still be corrected later on by means
of the overlap formula, whereas no such tool is available to correct
the scaling behavior. Hence it is important to optimize {\em scaling}
--- rather than chirality --- from the beginning.

Moreover, if one constructs a truncated approximate
GW fermions in a short range, there is no direct control over the level of locality
of the full GW fermion that one approximates.
Actually it is not even certain if
the GW solution that one would obtain by gradually extending the
range to infinity is local at all, since also non-local solutions
of the GWR exist. A simple example for that can be constructed
from the SLAC fermion \cite{SLAC}. The Dirac operator 
$D = D_{SLAC} (1+RD_{SLAC})^{-1}$ solves the GWR,
but it is non-local.}
The free spectrum of the {\em ``scaling optimal hypercube fermion''} (SO-HF)
is again close to a unit circle, on the same level as the 4d
spectrum in Fig.\ \ref{spec4d}.
There are also other ways to see that we are in the vicinity of
a GW fermion, for instance by summing over its 
violation (squared) in each site \cite{WBIH}, 
or by inserting the HF into the GWR and solving for $R$:
for the SO-HF, $R_{x,y}$ decays about 6 times faster
than the corresponding (pseudo-)$R$ for $D_{W}$.
If we compare the truncated perfect HF to $D_{W}$,
then this factor even amounts to $\approx 71$ in $d=2$, 
and to $\approx 75$ in $d=4$.

We now proceed to the 2-flavor Schwinger model, and we gauge the SO-HF
by attaching the couplings to the shortest lattice paths only.
When there are several shortest paths, the coupling is
split and attached to them in equal parts. Moreover we add
a clover term with coefficient 1. For the gauge part
we use the standard plaquette action (which is perfect for 2d
pure $U(1)$ gauge theory \cite{BWNP}).

\begin{figure}[htb]
\vspace{-7mm}
\begin{center}
\vspace{-5mm}
\hspace{14mm}
\def\fpsangle{270}
\epsfxsize=85mm
\fpsbox{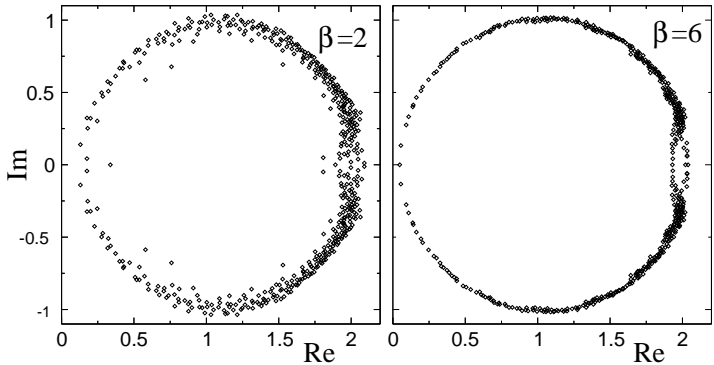}
\end{center}
\vspace{-32mm}
\caption{\it{The spectra of the 2d scaling optimal hypercube fermion (SO-HF)
for typical configurations at strong resp.\ weak coupling,
approximating a GW circle.}}
\label{spec2d}
\vspace{-3mm}
\end{figure}

\begin{figure}[hbt]
\vspace{-3mm}
\begin{tabular}{cc}
\hspace{-4mm}
\def\fpsangle{0}
\epsfxsize=60mm
\fpsbox{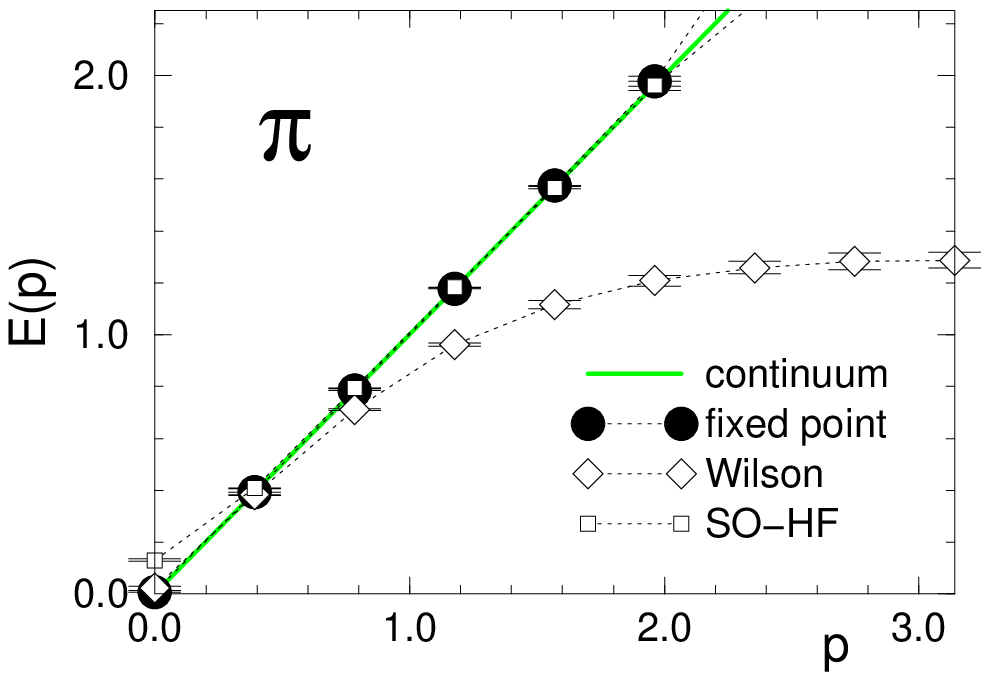}
\def\fpsangle{0}
\epsfxsize=60mm
\fpsbox{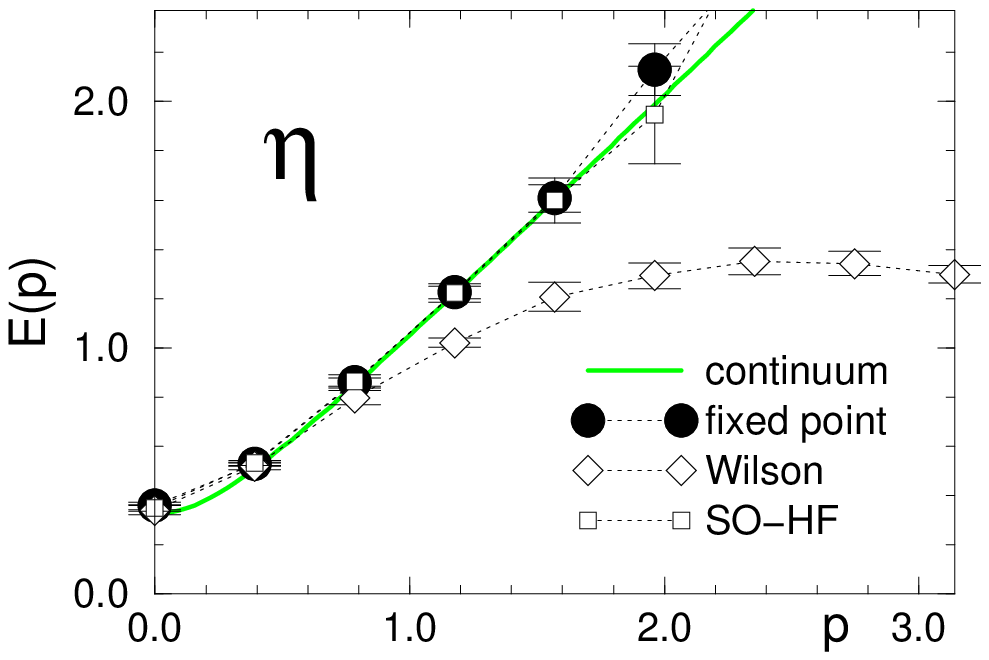}
\end{tabular}
\vspace{-3mm}
\caption{\it{The ``meson'' dispersion relations for some types of
2d fermions in the 2-flavor Schwinger model at $\beta = 6$.}}
\vspace{-4mm}
\label{meso-disp}
\end{figure}

This simple ``gauging by hand'' causes a further deviation from the GWR, 
which is increasingly manifest if the gauge background becomes rougher.
In Fig.\ \ref{spec2d} we show the spectra for typical configurations
at $\beta =2$ and at $\beta = 6$.
It turns out, however, that the SO-HF is indeed an approximate GW fermion
up to a considerable couplings strength. Next we test the scaling
behavior in the presence of gauge interaction. Our simulation results
(here and below) were obtained in collaboration with I.\ Hip.
They are based on 5000 quenched configurations on a
$16\times 16$ lattice at $\beta = 6$
(we use the same set of configurations for all types of fermions),
but the evaluation does include the fermion determinant, following
Ref.\ \cite{FHL}. Fig.\ \ref{meso-disp} shows the
dispersion relations for the two meson-type states, 
a massless triplet and one massive mode \cite{SaWi}, which we denote
as $\pi$ and $\eta$ (by analogy). Again the SO-HF is drastically
improved over the Wilson fermion (at $\kappa_{c}=0.25927$, from Ref.\
\cite{HLT}). It even reaches the same level as a (very mildly truncated) 
classically perfect action \cite{LP}, which was 
parameterized by 123 independent couplings per site, whereas only
6 such couplings are used for the SO-HF. Therefore, it is realistic
to extend the HF formulation to QCD, and in fact it has been shown 
already that minimally gauged 4d HFs can indeed be applied in QCD 
simulations \cite{4dsim,SESAM}. However, in QCD as well as in the Schwinger model,
we observed as an unpleasant feature of the directly applied HF
a strong additive mass renormalization. Using the SO-HF,
even at $\beta =6$ the $\pi$ mass is renormalized from 0 to 0.13, 
as Fig.\ \ref{meso-disp} shows. 
This corresponds to a lowest real eigenvalue around 0.03,
cf.\ Fig.\ \ref{spec2d}.
As an even more striking example, the 4d massless HF, minimally
gauged by hand and applied to QCD at $\beta =5$, leads to a ``pion mass'' 
of 3.0 \cite{BBCW}, and at $\beta =6$ the critical bare HF mass amounts 
to $-0.92$ \cite{SESAM} (which is unfavorable for the level of locality
before truncation, and hence for the magnitude of the truncation effects).
We overcome this problem in the next section
by inserting the SO-HF into the overlap formula.

[As a further alternative to negative bare mass and overlap, we can
reach the exact chiral limit solely by the use of fat links \cite{WBIH}.
This is essentially equivalent to gauging the HF such that the interacting
GWR is just violated modestly. As yet another possibility one may attach an 
amplification factor $\gsim 1$ to each link \cite{Eick-prep}.]

\section{Improved overlap fermions}

We now perform the second step in our program and insert the
SO-HF --- which is an approximate GW fermion with an excellent
scaling behavior ---
into the overlap formula (\ref{overlap}) with $R = R^{(st)}$
given in eq.\ (\ref{Rst}).
This leads to an exact GW fermion and therefore to all the
nice properties related to chirality, which are extensively discussed
in the recent literature on GW fermions: correct anomalies;
no renormalization of mass zero, vector current and flavor non-singlet
axial vector current; no mixing of weak matrix elements;
no exceptional configurations \cite{Has};
correctly reproduced chiral symmetry breaking \cite{SCKY} etc.

\begin{figure}[hbt]
\begin{tabular}{cc}
\hspace{-4mm}
\def\fpsangle{270}
\epsfxsize=42mm
\fpsbox{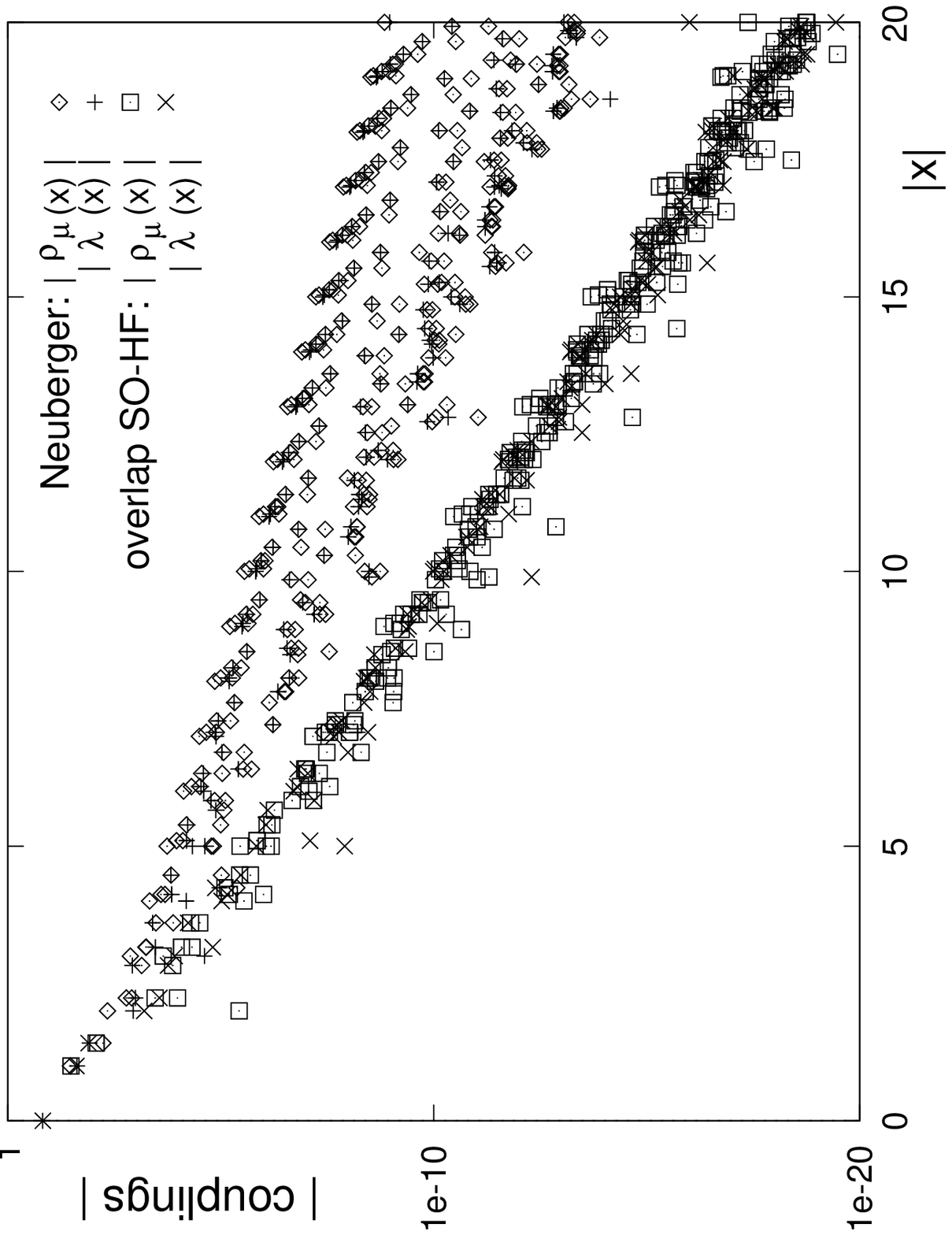}
\def\fpsangle{0}
\epsfxsize=66mm
\fpsbox{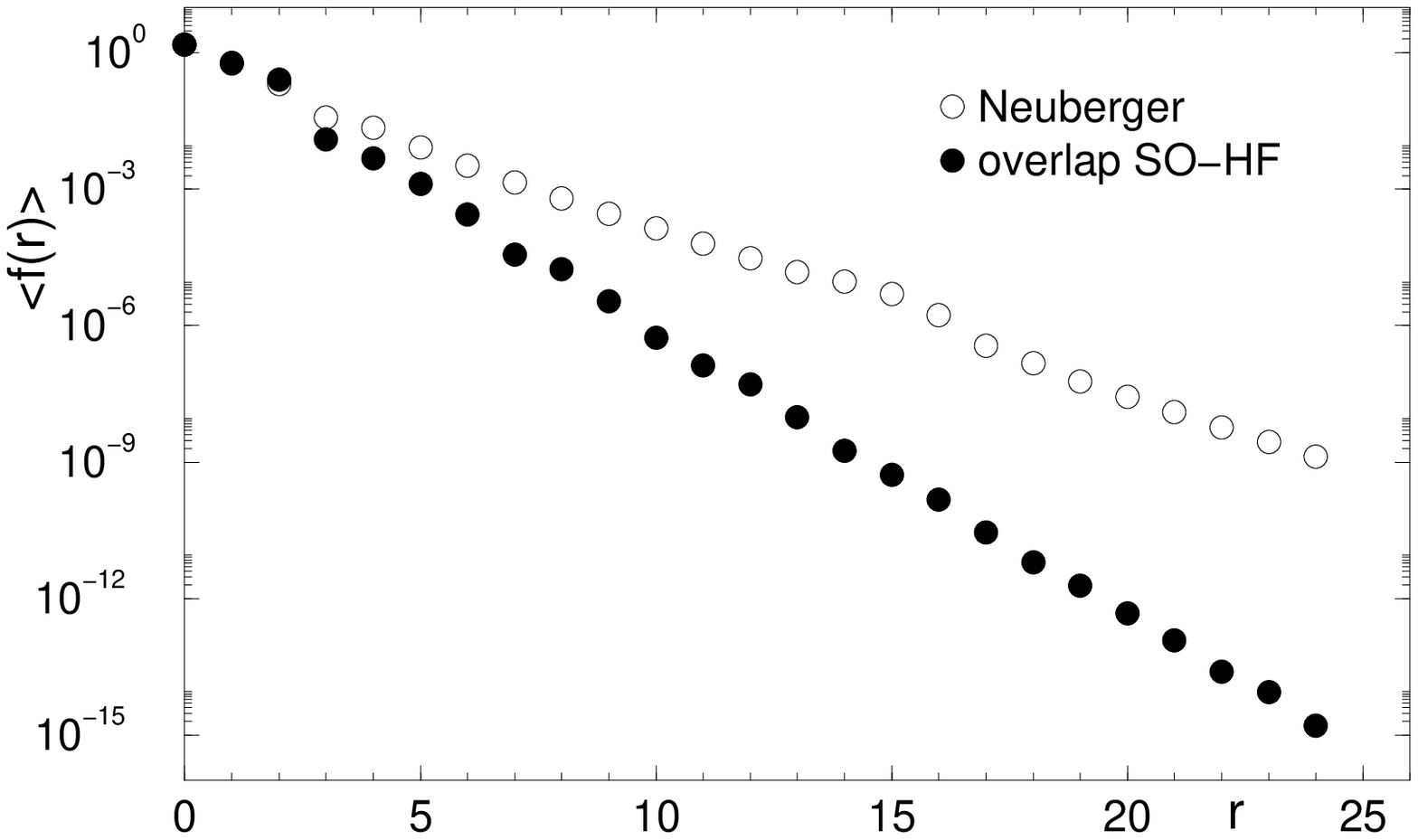}
\end{tabular}
\vspace{-4mm}
\caption{\it{The level of locality for the Neuberger fermions vs.\
overlap SO-HF: the decay of the free couplings in all directions
in infinite volume (left) and the decay of the ``maximal correlation''
over a distance $r$, as defined in Ref.\ \protect\cite{HJL}, 
on a $24 \times 24$ lattice at $\beta =6$ (right).
The figure on the left also illustrates how well
rotation invariance is approximated.}}
\vspace{-3mm}
\label{local}
\end{figure}

Our first prediction was that the level of locality should be 
improved over the Neuberger fermion, and this is clearly confirmed,
both, in the free and in the interacting case. 
Fig.\ \ref{local} compares the
decay of the couplings of the free fermion (left) and the
decay of the ``maximal correlation'' $f$ over a certain distance
$r$ --- as suggested in Ref.\ \cite{HJL} --- at $\beta =6$ (right).
\footnote{One puts a unit source at some site $x$ and defines
$f(r) :=  \ ^{\rm max}_{~~y} \ \{ \ \Vert \psi (y) \Vert 
\ \rule[-1.3mm]{0.4mm}{4mm}  \ \vert x-y \vert = r \ \}$.}
(In the latter figure one could still try to improve the locality
in both cases by deviating from $R^{(st)}$ and tuning $R$ for optimal
locality.) 
Next we want to verify if the good scaling quality survives
the modification due to the overlap formula. For the free
fermions, this is confirmed in an impressive way, see
Fig.\ \ref{freescale}.
\footnote{The dispersion curves of the overlap fermions just stop
inside the Brillouin zone due to the square root.
The end-points can be shifted by choosing $R_{x,y} = \delta_{x,y}/(2\mu )$
and varying $\mu$. We show $\mu = 1$ which is a reasonable
choice; if we push the end-point to far, the danger of doubling
increases. On the other hand, the regime of small momenta is needed, of course.} 
On fine lattices,
the scaling is practically identical before and after the use of 
the overlap formula; only as the lattice becomes really coarse, the overlap
does some harm to the scaling quality at some point.

\begin{figure}[hbt]
\begin{tabular}{cc}
\hspace{-4mm}
\def\fpsangle{0}
\epsfxsize=60mm
\fpsbox{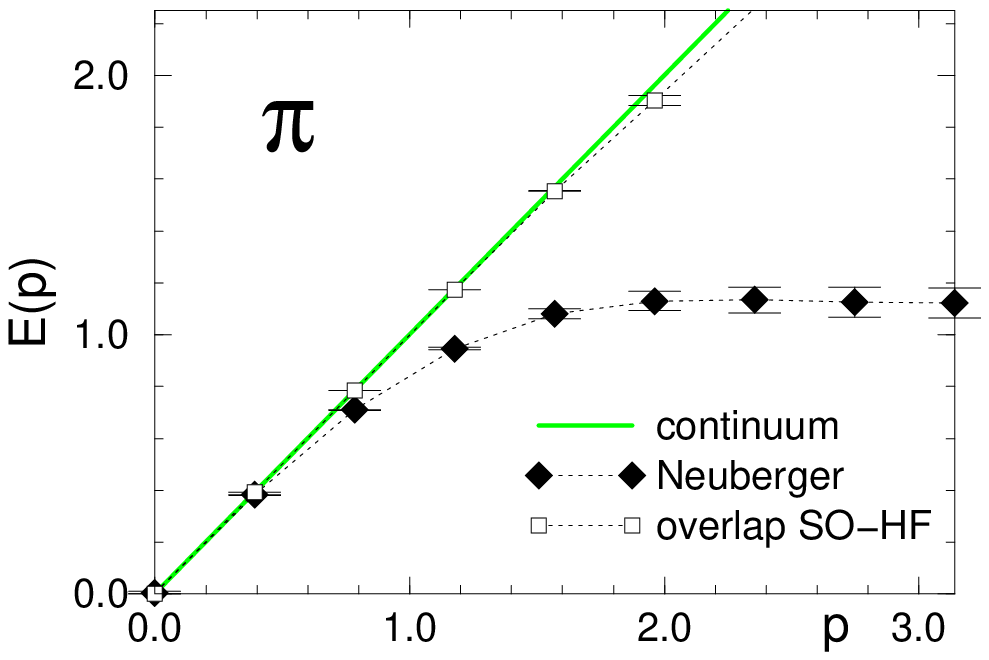}
\def\fpsangle{0}
\epsfxsize=60mm
\fpsbox{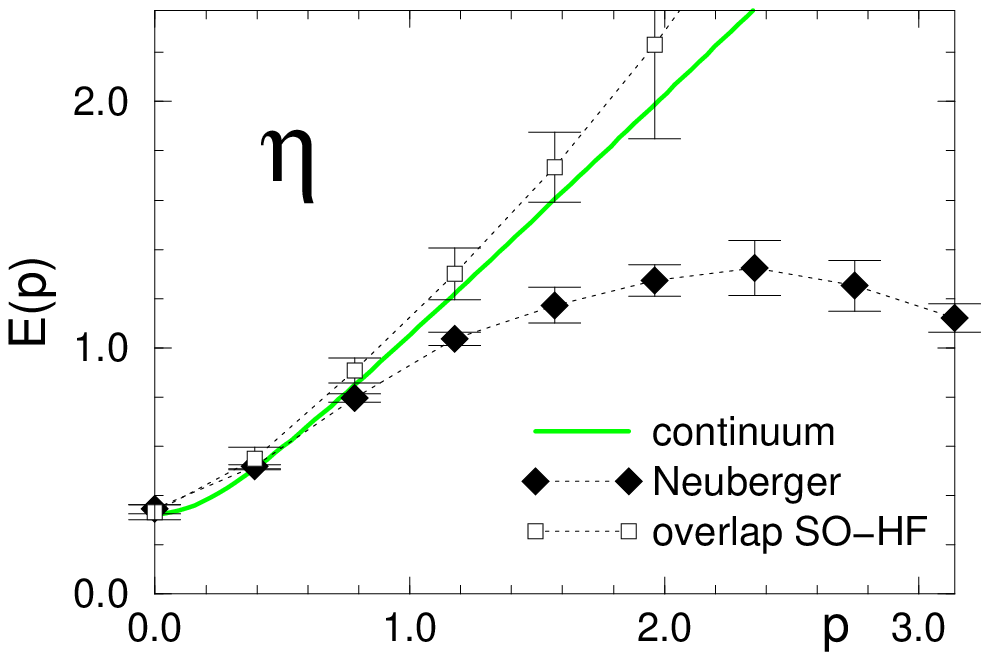}
\end{tabular}
\vspace{-5mm}
\caption{\it{The ``meson'' dispersion relations for two types of overlap
fermions at $\beta = 6$.}}
\vspace{-4mm}
\label{meso-disp-ov}
\end{figure}

In the presence of gauge interaction,
we consider again the ``meson'' dispersions, and we observe again the persistence
of the improvement, see Fig.\ \ref{meso-disp-ov}.

Furthermore, continuous rotation invariance is approximated 
much better for the HF than
for the Wilson fermion (after all, also this holds {\em exactly} 
for the perfect fermion \cite{BWNP}). 
In Ref.\ \cite{WBIH} this was also tested in the 
interacting case by measuring how smoothly correlations decay
with the Euclidean distance, and we observed that the SO-HF is again
by far superior over the Wilson fermion; once more it reaches
the same level as the classically perfect action.
The overlap formula (\ref{overlap}) suggests that this property is essentially
inherited for the overlap fermions, and indeed we observed a
similar level for the Wilson fermion and the Neuberger fermion on
one hand, and for the SO-HF and the overlap SO-HF on the other hand.
This confirms that also the strongly improved approximate rotational invariance
of the SO-HF survives if it is turned into an overlap fermion.
For the free fermion, this progress can also been observed in Fig.\ 
\ref{local} (left) from the width of the ``cones''.

\subsection{Chiral correction in terms of a power series}

In $d=2$ we can afford an exact evaluation of the notorious
square root in the overlap formula, but in QCD this is not feasible
any more. In the recent literature, a number of iterative
procedures have been suggested for the Neuberger fermion \cite{HJL,algos}.
In Ref.\ \cite{WBIH} we presented a new method, which is very
simple and robust, and which is especially designed for the
case where $D_{0}$ is an approximate GW fermion already.
We evaluate the square root in
\begin{equation}
D = 1 + \frac{A}{\sqrt{A^{\dagger}A}} \ , \quad A := D_{0} - 1
\end{equation}
as a power series in $\varepsilon := A^{\dagger}A -1$.
For $D_{0}=D_{HF}$ we have $\Vert \varepsilon \Vert \ll 1$
(if the configuration is not extremely rough),
hence the expansion converges rapidly. On the other hand,
for the case of the Neuberger fermion, i.e.\ for $D_{0}=D_{W}$, 
this expansion fails to converge even in the free case, 
which is presumably the reason why
it had not been considered in the earlier literature.
We call this method a ``perturbative chiral correction'',
where the perturbative expansion refers to the GWR violation
$\varepsilon$ (and not to the coupling $g$). It yields a
Dirac operator of the form
\begin{equation}
D_{p\chi c} = 1 +A Y \ .
\end{equation}
For the chiral correction to $O(\varepsilon^{n})$, $Y$ is a polynomial
in $A^{\dagger}A$ of order $n$, for instance,
\begin{eqnarray}
Y &=& [3 - A^{\dagger}A]/2 \qquad  {\rm for} \quad n=1 \ , \\
Y &=& [15 - 10A^{\dagger}A + 3 (A^{\dagger}A)^{2}]/8
\qquad {\rm for} \quad n=2,\ {\rm etc.} \nonumber
\end{eqnarray}
Hence the computational effort
amounts roughly to $1+2n$ matrix-vector multiplication (the matrix being
$A$ or $A^{\dagger}$), i.e.\ it increases {\em only} linearly (though
the convergence is also just linear).
Actually this represents a fermion with couplings of range
$1+2n$ in each component, and it would be very tedious to implement
it explicitly, even for $n=1$. However, due to
its specific form we never need to do so; $D_{p\chi c}$ can always
be evaluated by iteration of the above matrix-vector products.

\begin{figure}[hbt]
\begin{tabular}{cc}
\hspace{-5mm}
\def\fpsangle{270}
\epsfxsize=45mm
\fpsbox{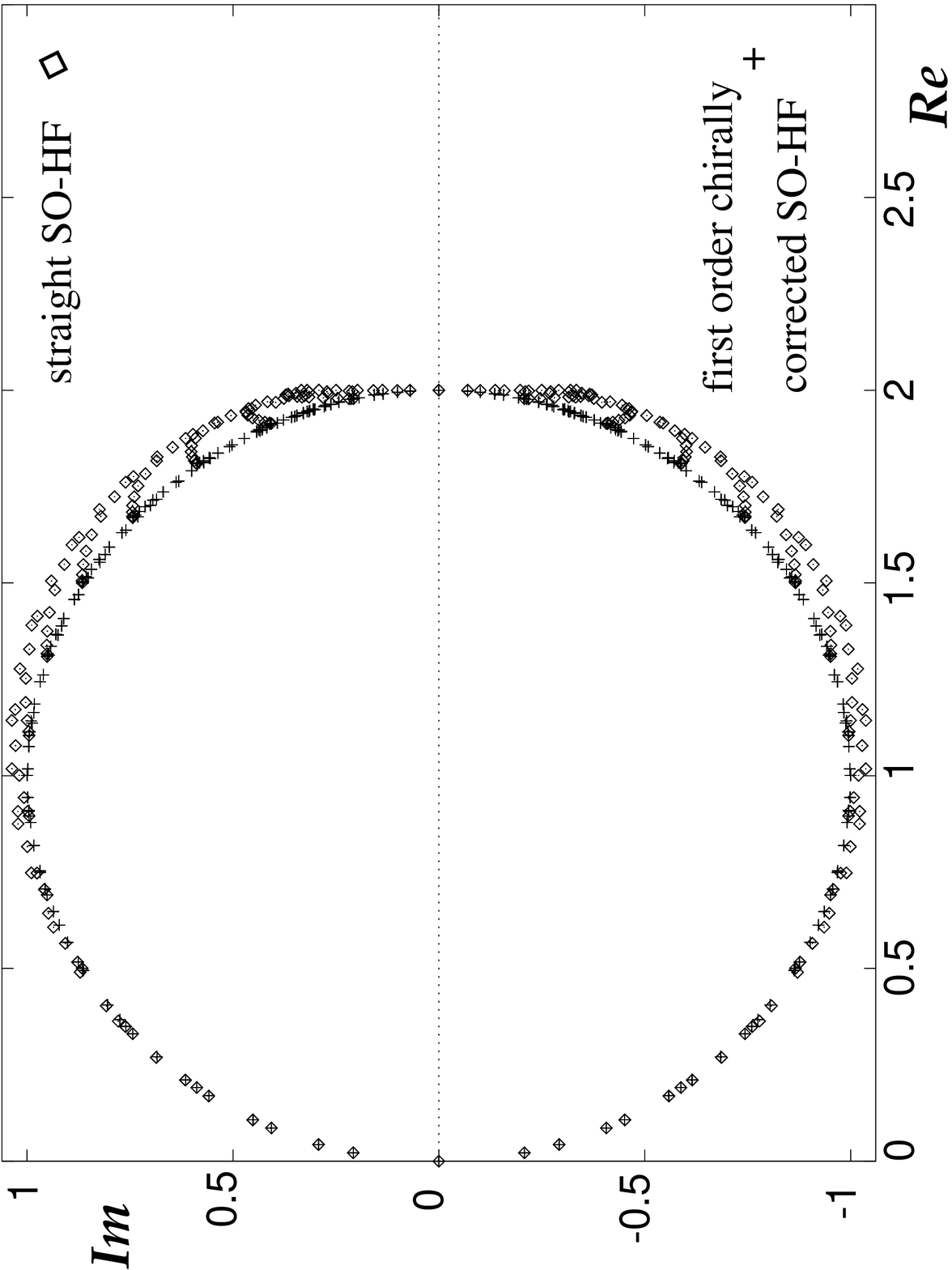}
\hspace{1mm}
\def\fpsangle{0}
\epsfxsize=60mm
\fpsbox{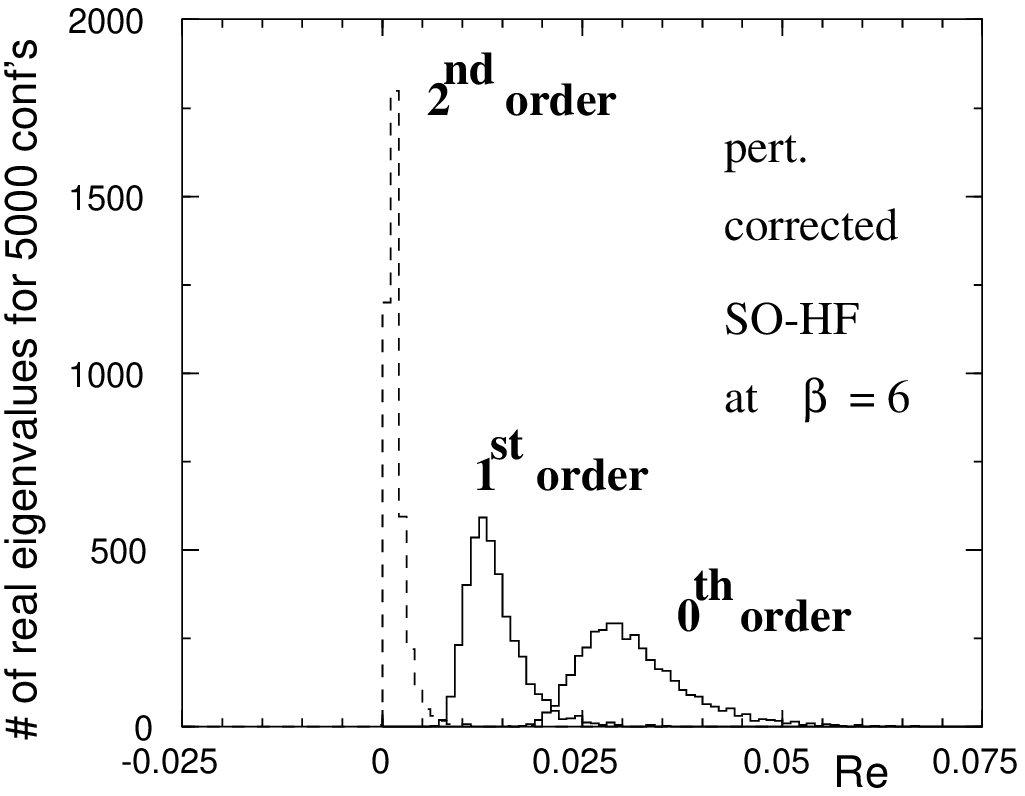}
\end{tabular}
\vspace{-3mm}
\caption{\it{The convergence of the SO-HF to a GW fermion
under perturbative chiral correction (on a $16 \times 16$
lattice): the free fermion spectrum (left) and the
distribution of the small real eigenvalues at $\beta =6$,
moving rapidly closer to 0 (right).}}
\vspace{-3mm}
\label{chir-cor}
\end{figure}

The crucial question now is if the first few orders are sufficient
already to do most of the chiral correction. We first look at the
free SO-HF, and Fig.\ \ref{chir-cor} (left) shows that the first order
alone does practically the full job. In this context, we also obtain
a geometric picture of the effect of the overlap formula: for
$R_{x,y} = \delta_{x,y}/(2\mu )$ it can be viewed as a {\em projection}
of the eigenvalues onto the circle with center and radius $\mu$.
This projection is often close to radial.

In the interacting case, the convergence to the circle under iteration is slowest
in the arc around 0. As an example, we show in Fig.\ \ref{chir-cor}
(right) a histogram of the small real eigenvalues at $\beta =6$.
We see that the mass renormalization is removed almost completely if we
proceed to $O(\varepsilon^{2})$. 

\subsection{Behavior in extremely rough gauge configurations}

For smooth configurations, it is easy to find a parameter $\mu$ ---
and hence a GW circle where the spectrum is mapped on --- such
that the small (large) real eigenvalues are mapped on 0 ($2\mu$).
This leaves the index unchanged (though it is defined by
exact zero modes now) and it provides a sensible
definition of the topological charge via the index theorem \cite{Has,ML}.
It also means that the doubling problem is safely avoided
for all typical configurations at moderate or large $\beta$.

\begin{figure}[hbt]
\begin{tabular}{cc}
\hspace{-6mm}
\def\fpsangle{0}
\epsfxsize=62.5mm
\fpsbox{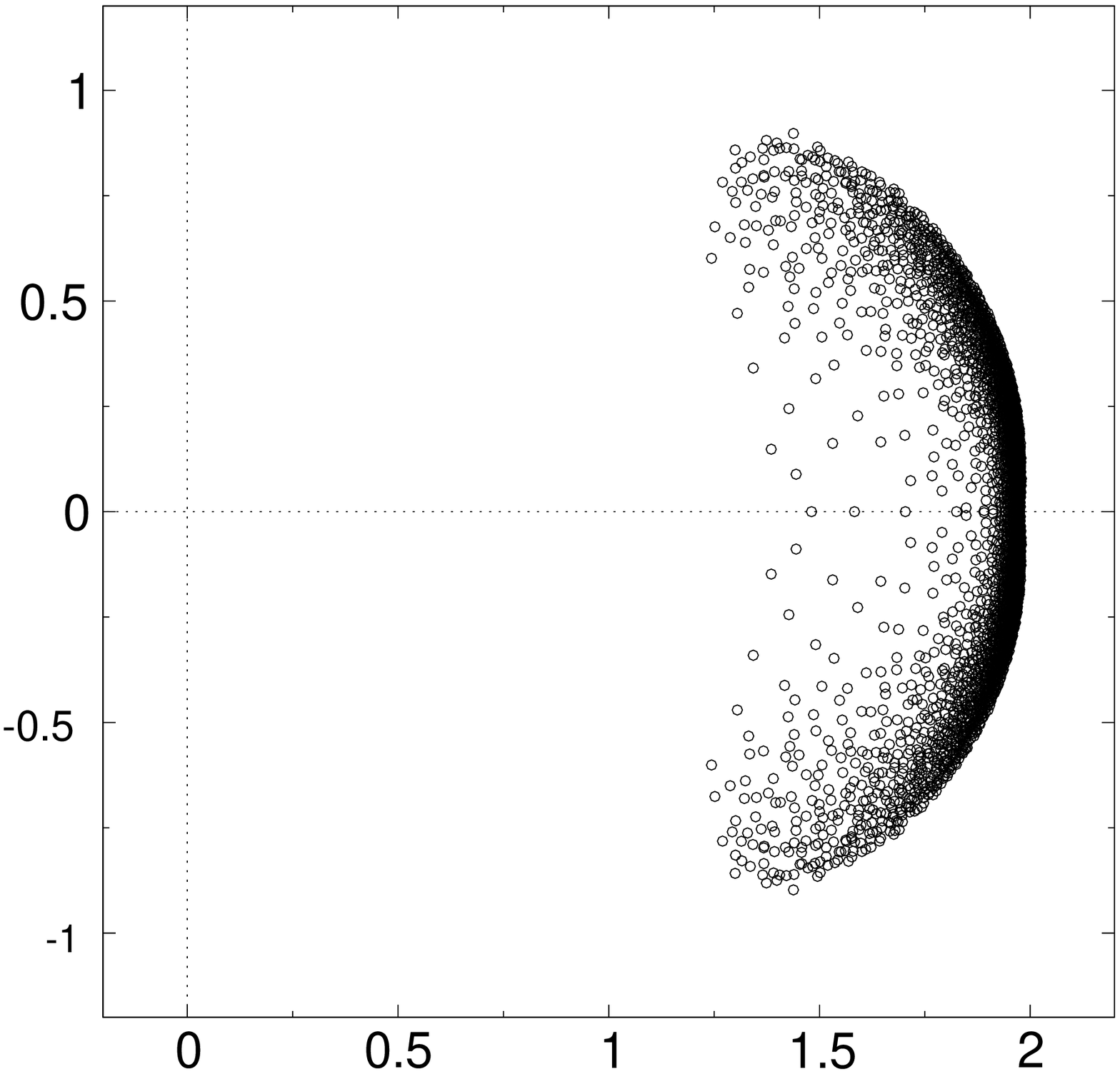}
\def\fpsangle{0}
\epsfxsize=60mm
\fpsbox{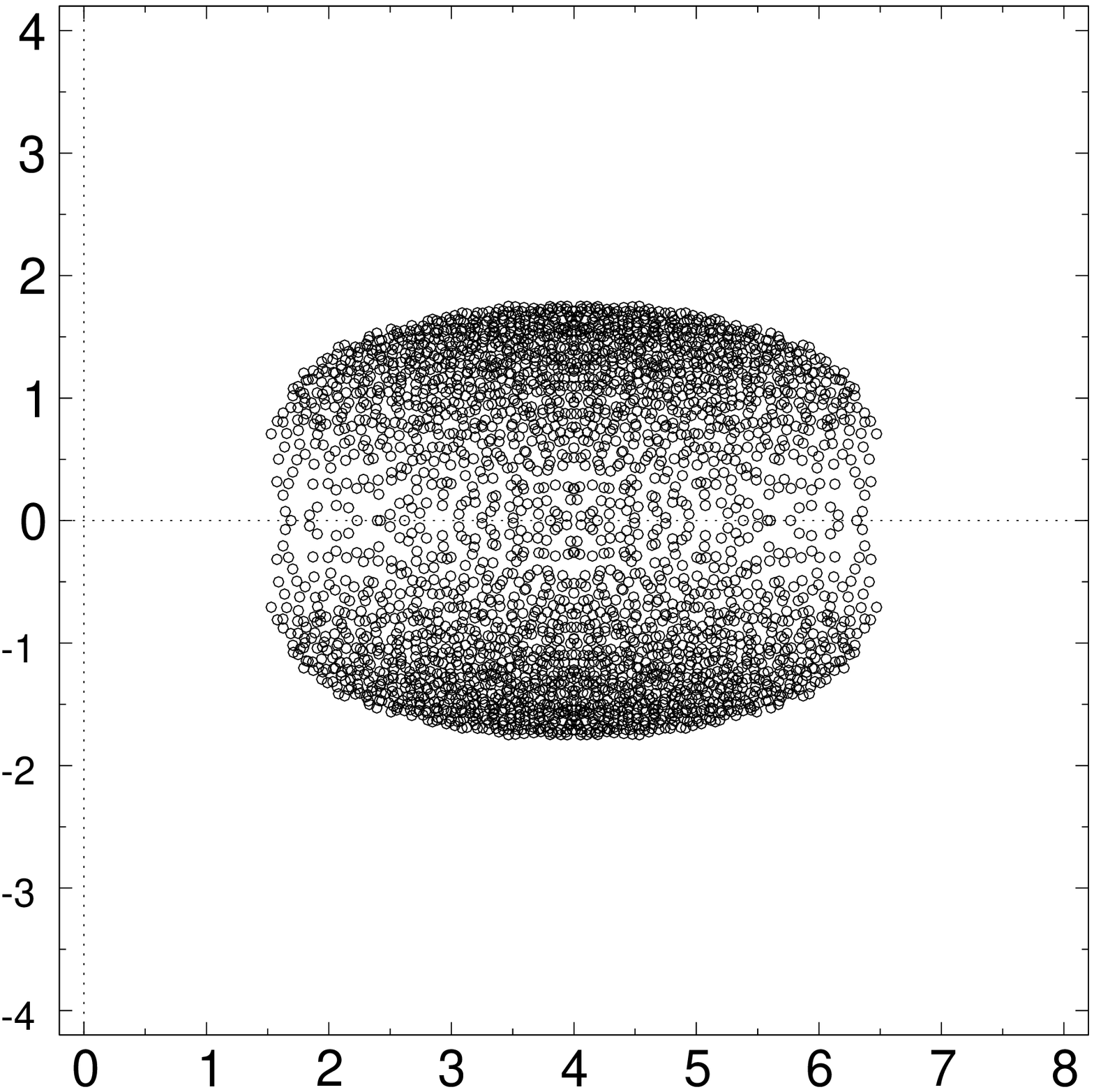}
\end{tabular}
\vspace{-1mm}
\caption{\it{The fermion spectrum for an extremely rough QCD test
configuration on a $4^{4}$ lattice: for the hypercube fermion
of Ref.\ \protect\cite{BBCW} (left), and for the Wilson fermion (right).}}
\vspace{-4mm}
\label{specQCD}
\end{figure}

However, if we dare proceeding to extremely strong coupling,
then it is not possible any more to find such a center $\mu$
of a GW circle, which does the right mapping for all typical
configurations. In fact, for extremely rough QCD test configurations
(on very small lattices) it was observed explicitly that {\em all} the
eigenvalues of the minimally gauged $D_{HF}$ are close to the arc of 
the GW circle, which is opposite to 0 \cite{IH4d}.
For $D_{W}$ the eigenvalues are densely scattered over
a wide area with a large real part. Examples for such spectra
of an extremely rough configuration are shown in Fig.\ \ref{specQCD},
which was provided by N.\ Eicker, I.\ Hip and Th.\ Lippert.
At a coupling strength where such configurations are frequent,
the doubling problem is back for those overlap fermions,
which are constructed from some simple $D_{0}$
(for the Schwinger model, this problem sets in around $\beta \approx 1$
\cite{IH4d}).
This agrees with the result of a strong coupling expansion 
(in the Hamiltonian formulation) which applies to $D_{Ne}$ \cite{BS}.
\footnote{For a Euclidean strong coupling hopping parameter expansion of $D_{Ne}$,
see Ref.\ \cite{IN}.}
In such cases, the construction of $D_{Ne}$ maps all (almost)
real eigenvalues onto the arc close to 2, hence additive mass
renormalization is back as well --- again in agreement with Ref.\ \cite{BS}.
In view of the latter point one could be tempted to just use
a large mass parameter $\mu >1$; this does not help, however, with respect
to the doubling problem.
At very strong coupling, only the (classically) perfect action would help.

As we mentioned earlier, also locality is in danger in that
regime \cite{HJL}, and we should therefore keep away from it.
As one more advantage of choosing $D_{0}$ to be an approximate GW
fermion, the regime of $\beta$ where we are (statistically)
on safe grounds is enlarged compared to the Neuberger fermion.\\

Of course, in the safe regime where $\beta$ is large enough
(in QCD this includes $\beta =6$ for sure \cite{HJL})
the chiral correction of $D_{0}=D_{HF}$ can also be carried out
by iteration methods different from the one described in the previous
Subsection, for alternative experiments see Ref.\ \cite{GWgames}. 
The efficiency in QCD is still to be compared, but for sure in any 
method the convergence will be much faster for $D_{0}=D_{HF}$ than 
for $D_{0}=D_{W}$.

\section{Conclusions}

Our program outlined in Sec. 2.1 has been realized in the
Schwinger model, and the properties of a resulting improved
overlap fermion have been tested extensively. They are all clearly
superior over the Neuberger fermion, confirming our prediction:
the overlap SO-HF scales much better, it is more local
and it comes much closer to rotation invariance.

The question now is the applicability of this program in $d=4$.
The 4d HF formulation is worked out already,
and the corresponding improved overlap fermion
is currently under investigation in QCD by
the SESAM collaboration in J\"{u}lich and Wuppertal.\\


{\bf Acknowledgment} Most of the results presented here are based on
my collaboration with Ivan Hip, and I would like to thank him for his 
crucial contributions. Furthermore I am indebted to P.\ Damgaard,
T.\ DeGrand, P. Hern\'{a}ndez, A.\ Hoferichter, K.\ Jansen, J.\ Jers\'{a}k,
F.\ Klinkhamer, Th.\ Lippert, K.F.\ Liu, M.\ L\"{u}scher, J.\ Nishimura,
P.\ Rakow, K.\ Schilling  and A.\ Thimm for useful comments. 
Finally I would like to thank  the organizers of this work-shop, 
in particular V.\ Mitrjushkin, for their kind hospitality in beautiful Dubna.


{\small

}


\begin{thebibliography}{99}

\bibitem{NN} H.\ Nielsen and M.\ Ninomiya, {\it Nucl.\ Phys.}\ {\bf B185} (1981) 20.
\bibitem{ML} M.\ L\"{u}scher, {\it Phys.\ Lett.\ } {\bf B428} (1998) 342.
\bibitem{GW} P.\ Ginsparg and K.\ Wilson, {\it Phys.\ Rev.\ } {\bf D25} (1982) 2649.
\bibitem{Kerl} W.\ Kerler, {\bf hep-lat/9905010}.
\bibitem{Has} P.\ Hasenfratz, V.\ Laliena and F.\ Niedermayer,
{\it Phys.\ Lett.\ } {\bf B427} (1998) 125.\\
P.\ Hasenfratz, {\it Nucl.\ Phys.\ } {\bf B525} (1998) 401.
\bibitem{Fuji} K.\ Fujikawa, {\it Nucl. Phys.} {\bf B546} (1999) 480. 
\bibitem{EPJC} W.\ Bietenholz, {\it Eur. Phys. J.} {\bf C6} (1999) 537.
\bibitem{Neu} H.\ Neuberger, {\it Phys.\ Lett.\ } {\bf B417} (1998) 141; 
{\bf B427} (1998) 353.
\bibitem{Rebbi} C.\ Rebbi, {\it Phys. Lett.} {\bf B186} (1987) 200.
\bibitem{CZ} T.-W.\ Chiu and S.\ Zenkin, {\it Phys. Rev.} {\bf D59} (1999) 074501.
\bibitem{UJW} U.-J.\ Wiese, {\it Phys.\ Lett.\ } {\bf B315} (1993) 417. \\
W.\ Bietenholz and U.-J.\ Wiese, {\it Nucl. Phys. B (Proc.\ Suppl.)}
{\bf 34} (1994) 516.
\bibitem{Melb} W.\ Bietenholz and U.-J.\ Wiese, 
{\it Nucl.\ Phys.\ B (Proc.\ Suppl.)} {\bf 47} (1996) 575.
\bibitem{BWNP} W.\ Bietenholz and U.-J.\ Wiese, {\it Nucl.\ Phys.\ } {\bf B464}
(1996) 319.
\bibitem{SUSY} W.\ Bietenholz, {\it Mod.\ Phys.\ Lett.\ } {\bf A14} (1999) 51.
\bibitem{Axel} A.\ Thimm, talks presented at LATTICE'99 and at this workshop.
\bibitem{Peli} G.\ Bodwin and E.\ Kovacs, {\it Phys. Lett.} {\bf B193} (1987) 283;
{\it Phys. Rev.} {\bf D35} (1987) 3198.
M.\ Campostrini, G.\ Curci and A.\ Pelissetto,
{\it Phys. Lett.} {\bf B193} (1987) 273.\\
A.\ Pelissetto, {\it Ann.\ Phys.\ } {\bf 182} (1988) 177.
\bibitem{BWPL} W.\ Bietenholz and U.-J.\ Wiese, {\it Phys.\ Lett.\ } {\bf B378} 
(1996) 222.
\bibitem{Capi} S.\ Capitani et al., {\it Phys.\ Lett.\ } {\bf B468} (1999) 150.
\bibitem{HJL} P.\ Hern\'{a}ndez, K.\ Jansen and M.\ L\"{u}scher, 
{\it Nucl.\ Phys.\ } {\bf B552} (1999) 363.
\bibitem{Neuloc} H.\ Neuberger, {\bf hep-lat/9911004}.
\bibitem{IH} I.\ Horv\'{a}th, {\it Phys.\ Rev.\ Lett.\ } {\bf 81} (1998) 4063.
\bibitem{ultraloc} W.\ Bietenholz, {\bf hep-lat/9901005}.
\bibitem{FN} F.\ Niedermayer, {\it Nucl.\ Phys.\ (Proc.\ Suppl.)} {\bf 73}
(1999) 105.
\bibitem{KFL} K.F.\ Liu, S.J.\ Dong, F.X.\ Lee and J.B.\ Zhang, {\bf hep-lat/9909061}.
\bibitem{JanWup}  P.\ Hern\'{a}ndez, K.\ Jansen and L.\ Lellouch, {\bf hep-lat/0001008}.
\bibitem{WBIH} W.\ Bietenholz and I.\ Hip, {\bf hep-lat/9902019}, 
to appear in {\it Nucl.\ Phys.\ B}; {\bf hep-lat/9908012}.
\bibitem{BBCW} W.\ Bietenholz, R.\ Brower, S.\ Chandrasekharan and U.-J.\ Wiese,
{\it Nucl.\ Phys.\ (Proc.\ Suppl.)} {\bf 53} (1997) 921.
\bibitem{AHF} F.\ Niedermayer, {\it Nucl.\ Phys.\ (Proc.\ Suppl.)} {\bf 53} (1997) 56.\\ 
W.\ Bietenholz, {\bf hep-lat/9911015}, to appear in {\it Int. J. Mod. Phys. A}.
\bibitem{chempot} W.\ Bietenholz and U.-J. Wiese, {\em Phys.\ Lett.\ } {\bf B426}
(1998) 114.
\bibitem{GatHip} C.\ Gattringer and I.\ Hip, {\bf hep-lat/0002002}.
\bibitem{SLAC} S.\ Drell, M.\ Weinstein and S. Yankielowicz,
{\em Phys. Rev.} {\bf D14} (1976) 487; 1627.
\bibitem{FHL} F.\ Farchioni, I.\ Hip and C.B.\ Lang,
{\it Phys. Lett.} {\bf B443} (1998) 214.
\bibitem{SaWi} I.\ Sachs and A.\ Wipf, {\it Helv. Phys. Acta} {\bf 65} (1992) 652.
\bibitem{HLT} I.\ Hip, C.B.\ Lang and R.\ Teppner,
{\it Nucl. Phys. B (Proc. Suppl.)} {\bf 63} (1998) 682.
\bibitem{LP} C.B.\ Lang and T.\ Pany, {\it Nucl.\ Phys.\ } {\bf B513} (1998) 645. 
\bibitem{4dsim}  K.\ Orginos et al., 
{\it Nucl.\ Phys.\ (Proc.\ Suppl.)} {\bf 63} (1998) 904.\\
T.\ DeGrand, {\it Phys.\ Rev.\ } {\bf D58} (1998) 094503.
\bibitem{SESAM} N.\ Eicker et al., {\it Comput.\ Phys.\ Commun.\ } {\bf 119} (1999) 1.
\bibitem{Eick-prep} N.\ Eicker et al., {\it in preparation}.
\bibitem{SCKY} S.\ Chandrasekharan, {\it Phys. Rev.} {\bf D60} (1999) 074503.\\
Y.\ Kikukawa and A.\ Yamada, {\it Nucl. Phys.} {\bf B547} (1999) 413. 
\bibitem{algos} H.\ Neuberger, {\it Phys. Rev. Lett.} {\bf 81} (1998) 4060.\\
A.\ Bori\c{c}i, {\it Phys. Lett.} {\bf B453} (1999) 46.\\
C.\ Liu, {\it Nucl. Phys.} {\bf B554} (1999) 313.
\bibitem{IH4d} I.\ Hip, {\it private communication}.
\bibitem{BS} R.\ Brower and B.\ Svetitsky, {\bf hep-lat/9912019}.
\bibitem{IN} I.\ Ichinose and K.\ Nagao, {\bf hep-lat/9910031}.
\bibitem{GWgames} T.\ DeGrand, {\bf hep-lat/9908037}.

\end{thebibliography}
\end{document}